\documentclass[structabstract]{aa} 

\usepackage{graphicx}
\usepackage{txfonts}
\begin{document}

   \title{Direct measurement of the distribution of dark matter with strongly lensed gravitational waves}

   \author{Shuo Cao\inst{1,2}
          \and
           Jingzhao Qi \inst{3}
           \and
           Zhoujian Cao\inst{1,2}
          \and
           Marek Biesiada\inst{4}
          \and
          Wei Cheng\inst{5}
          \and
           Zong-Hong Zhu\inst{1,2}\thanks{\email{zhuzh@bnu.edu.cn}}
          }

   \institute{Department of Astronomy, Beijing Normal University,
100875, Beijing, China;
\and
Advanced Institute of Natural Sciences, Beijing Normal
University at Zhuhai 519087, China;
         \and
         Department of Physics, College of Sciences,
Northeastern University, Shenyang 110004, China;
         \and
          National Centre for Nuclear Research,
Pasteura 7, 02-093 Warsaw, Poland;
         \and
         School of Science,
Chongqing University of Posts and Telecommunications, Chongqing
400065, China}

   \authorrunning{Cao et al.}
   \titlerunning{Direct measurement of the distribution of dark matter}

   \date{Received XXX/ Accepted XXX}

  \abstract
{In this Letter, we present a new idea of probing the distribution of
dark matter exhibiting elastic and velocity-independent self-interactions. These interactions might be revealed in multiple
measurements of strongly lensed gravitational waves, which can be
observationally explored to determine the strength of self-scatterings. Specifically, each individual galactic-scale strong-lensing system whose source is a coalescing compact binary emitting gravitational waves will provide a model-independent measurement of the shear viscosity of dark matter along the line of sight. These
individual measurements could be a probe of large-scale distribution
of dark matter and its properties. Our results indicate that with 10-1000
strongly lensed gravitational waves from ET and DECIGO, robust
constraints on the large-scale distribution of self-interacting dark
matter might be produced. More stringent limits on the dark matter
scattering cross-section per unit mass ($\sigma_{\chi}/m_{\chi}$) relevant to galaxy and cluster scales are also expected, compared with the conservative estimates obtained in the electromagnetic domain. Finally, we discuss the effectiveness of our method in the context of self-interacting dark matter particle physics.}


   \maketitle
%

\section{Introduction}

It is well known that dark matter (DM) -- the
dominant component of virialized objects (galaxies and clusters) --
is one of the largest open questions in modern astrophysics.
Especially the collisionless cold DM (CCDM), which is
widely recognized by the astrophysics community and strongly
supported by the current observations of large-scale structures
(e.g. galaxy clusters), follows the homogeneous and isotropic
distribution in the Universe \citep{Schaye15,Springel18}. However,
the CCDM paradigm still suffers from the well-known small-scale problem that
concerns some observed features of galaxies and dwarf galaxies
(e.g., missing satellite, core-cusp ,or too-big-to-fail problems) \citep{Read17}.
Interestingly, there are suggestions that self-interacting (SI) DM (SIDM) might successfully explain these discrepancies. This
provides an interesting alternative clue for the current cosmological
model \citep{Tulin18}.  Many ideas have been proposed to explore the
possibilities that DM SI might generate cosmic
accelerated expansion \citep{Zimdahl01,Cao11} and non-zero
cosmological shear viscosity \citep{Hawking1966}. More interestingly, SIDM with velocity-dependent SI has gathered
growing attention because it can provide a consistent fit
on the scattering cross sections per unit mass that is applicable to  different scales
\citep{Hayashi20}. Despite these advantages, the nature of
SIDM remains unknown. It is
in particular still debated whether SIDM traces CCDM in the Universe because no direct detection of its
distribution has been performed.

According to General Relativity, absorption and dispersion of
gravitational waves (GWs) could be neglected in a perfect-fluid
Universe \citep{Ehlers1996}. This theoretical point of view has been
widely applied in some recent works \citep{Qi19}. However, the
hypothesis of transparent GWs in the EM domain remains
untested experimentally \citep{Abbott16}. In this Letter, we propose
a new idea of measuring the properties of DM with elastic
and velocity-independent SI through the multiple
measurements of galactic-scale strong gravitational lensing systems
with neutron star (NS) mergers acting as background sources. The
advantage of our method is that I) viscously damped
gravitational waves from these standard sirens would reach the
observer, accompanied by the electromagnetic radiation in the form
of short and intense bursts of $\gamma$ rays \citep{Abbott17}. II)
Each individual GW - galaxy strong-lensing system will provide a
model-independent measurement of the fluid shear viscosity of dark
matter along the line of sight. Therefore, with a sample of
viscosity measurements at different positions on the sky, the distribution of dark matter in the Universe might be tested directly. III)
The strongly lensed NS-NS systems are ideal laboratories for deriving
robust constraints on the the scattering cross sections per unit
mass relevant for different scales, from ultra-dwarf galaxies to
relaxed galaxy clusters.

\section{Method}

The LIGO and Virgo Collaborations have so far released the measurements of luminosity distances for 90 GW sources \citep{Abbott21}. However, we stress that the luminosity distances $D_{L,obs}$ were derived from the observed strain $h(t)$
and frequency evolution $f(t)$ under the assumption that DM in
the Universe is treated as a perfect fluid. On the other hand, if
the DM can be better represented as a non-ideal fluid characterized
by a shear viscosity term $\eta$, then the relation
$\beta = \frac{16\pi G}{c^3} \eta$ between the GW damping rate ($\beta$)
and DM viscosity ($\eta$) is simple. This relation, which was originally
proposed in \citet{Hawking1966} with the $c=8\pi G =1$ convention, was also derived in the framework of geometric units \citep{Ehlers1996,Atreya}. In this analysis, we reintroduce the fundamental constants in the formulae. Considering the inverse relation between the amplitude of GW waveform and the luminosity distance, when the
GW damping rate is taken into account, the viscosity-free luminosity
distance ($D_L$) inferred from the standard siren GW signal is
modified to
\begin{eqnarray}\label{DLeff}
D_{L,eff}(z, \beta) = D_L(z) e^{\beta D(z)/2},
\end{eqnarray}
where $D_{L,eff}$ and $D$ represent the so-called effective
luminosity distance and comoving distance, respectively
\citep{Goswami2017,Cao21}. More recently, a method of measuring the
viscosity of DM in the cosmological context has been suggested
and implemented for the currently 11 GW events that were released by LIGO and
Virgo Collaborations \citep{Lu18}. However, a number of concerns were
also raised about the robustness of the above test: precise measurements of distances that are unaffected by
viscosity in the light of problems with host galaxy identification and
redshift determination for coalescing binary black holes (BH) are not available. This
motivates the search for other methods to probe the DM viscosity. In
this Letter we propose strongly lensed GWs
produced by coalescing NSs for this purpose.

\begin{figure}
\centering
\includegraphics[scale=0.65]{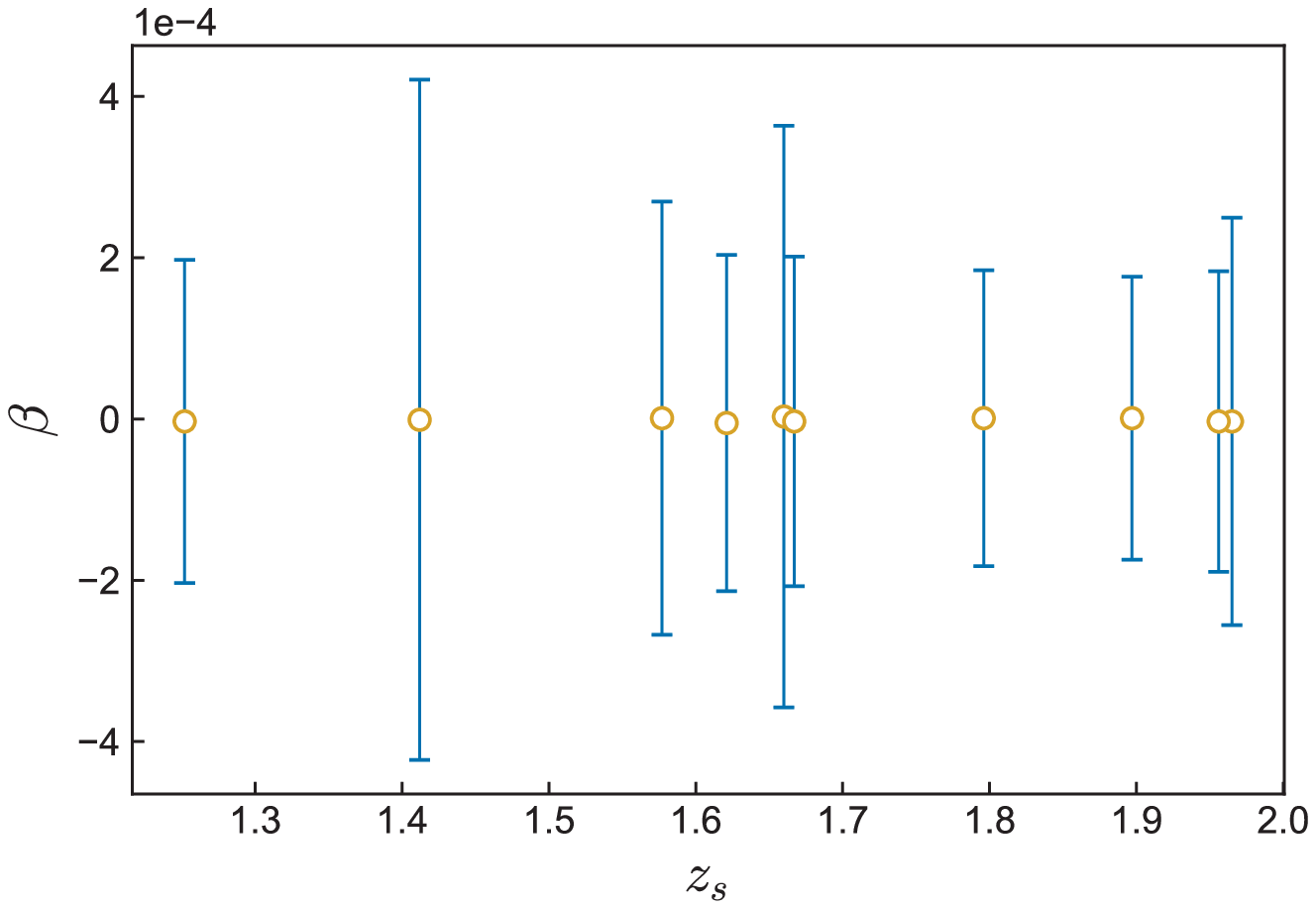}
\includegraphics[scale=0.65]{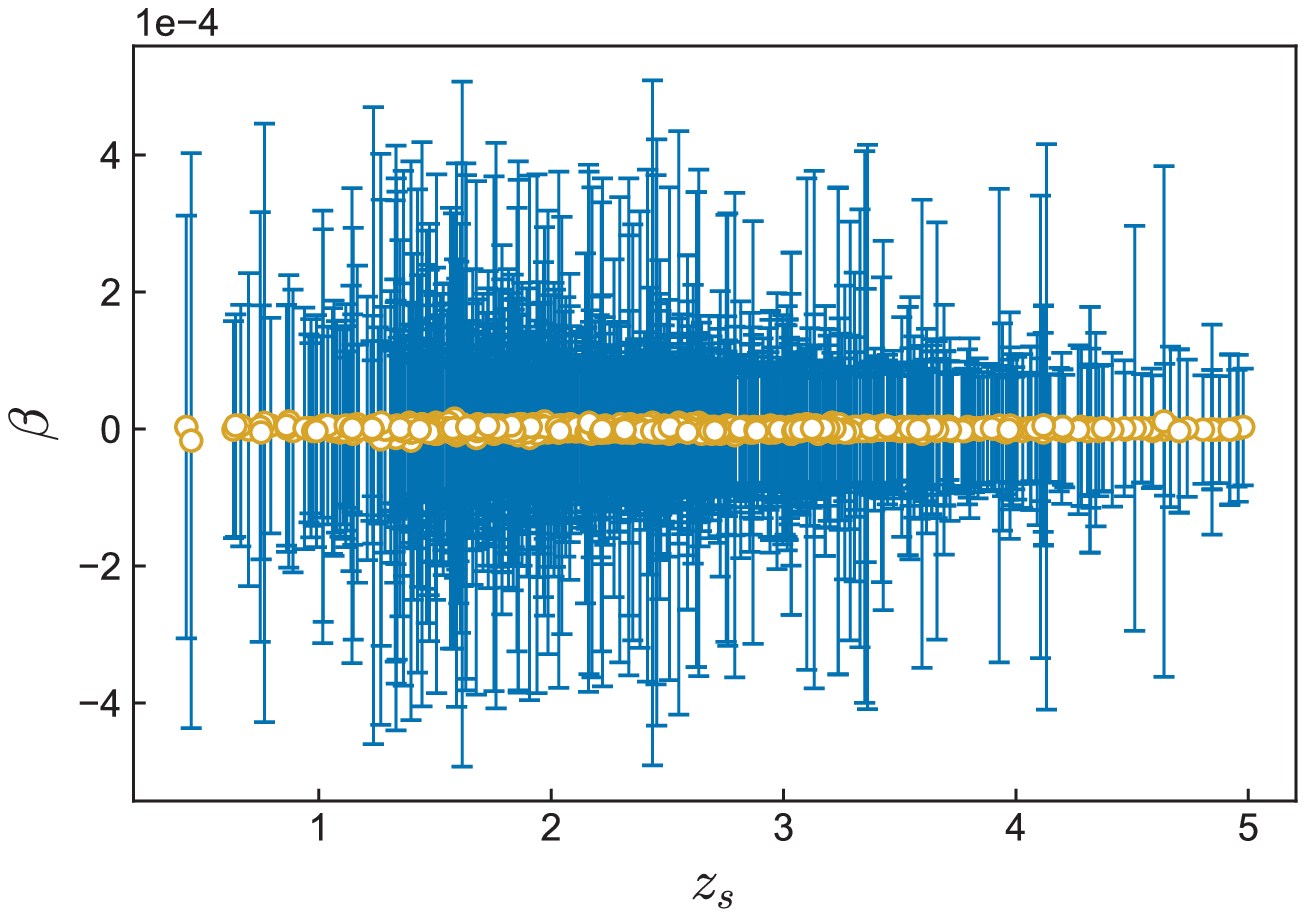}
\caption{Individual measurements of the damping rate of gravitational waves based on future observations of strongly lensed NS-NS systems from ET (upper panel) and DECIGO (lower panel).
}\label{fig1}
\end{figure}

As one of the successful predictions of General Relativity, strong
gravitational lensing by galaxies has become one of the most
important tools in studying cosmology \citep{Cao15} and the galaxy structure
and evolution \citep{Cao16}. With the dawn of GW
astronomy, the robust prediction suggested that a considerable
number of GW signals from inspiralling NSs would be
gravitationally lensed, focusing on the third generation of
ground-based GW detectors \citep{Ding15} and on the second generation of
space-based GW detectors \citep{Piorkowska21}, respectively. In this
Letter, our idea relies on the derived multiple distances in a
specific strong-lensing system (with the background GW source at
redshift $z_s$ and the lensing galaxy at redshift $z_l$), with
strongly lensed GW signals observed together with their
electromagnetic (EM) counterparts. We outline below three
approaches to measure the angular diameter distances between the
observer and the lens ($D_{A}(z_l)$), between the observer and the
source ($D_{A}(z_s)$), and between the lens and the source
($D_{A}(z_l,z_s)$).

Firstly, with the measurements of the central velocity dispersion and
the location of multiple images (Einstein radius) of a strong-lensing system, the ratio of two angular diameter distances
$D_{A}(z_l,z_s)/D_{A}(z_s)$ can be precisely assessed. Specifically, by
assuming a power-law model to describe the lens mass distribution
($\rho \sim r^{- \gamma}$), the lens potential relates to the
moments of the stellar distribution function through the Jeans
equation \citep{Koopmans05}. The combination of the lens mass and the
dynamical mass inside the Einstein radius ($\theta_E$) yields
\citep{Cao15}
\begin{equation} \label{Einstein} \frac{D_{A}(z_l,z_s)}{D_{A}(z_s)} =   \frac{\theta_E} {4 \pi}
\frac{c^2}{\sigma_{ap}^2}  \left( \frac{\theta_E}{\theta_{ap}}
\right)^{\gamma-2} f(\gamma)^{-1}
,\end{equation}
where the luminosity-averaged velocity dispersion $\sigma_{ap}$
is measured within the aperture $\theta_{ap}$ (see \citet{Cao15} for
the expression of $f(\gamma)$ in the form of the radial mass profile
slope). Therefore, the observations of $\sigma_{ap}$, $\theta_E$,
$\theta_{ap}$ , and $\gamma$ enables precise measurements of
the distance ratio $D_{A}(z_l,z_s)/D_{A}(z_s)$. Secondly, the time-delay
distance in a specific GW - galaxy strong-lensing system can be measured.  The
time-delay distance is defined as
\begin{equation}
D_{\mathrm{\Delta t}}(z_l, z_s) \equiv\frac{D_{A}(z_l)
D_{A}(z_s)}{D_{A}(z_l, z_s)}=\frac{c}{1+z_l}\frac{\Delta
t_{i,j}}{\Delta \phi_{i,j}}.
\end{equation}
It could be measured precisely and accurately for transient sources such as GW signals from coalescing binaries through the precise measurement of the time delay and the well-reconstructed Fermat potential difference
($\Delta\phi_{i,j}$) between multiple images, that is, the multiple GW signals. The EM counterpart
would allow the host galaxy and lens galaxy identification, and
dedicated lens modelling techniques would enable a precise
reconstruction of the lens mass distribution. More importantly, the time
difference ($\Delta t_{i,j}$) in the arrival times of two signals
(at angular coordinates $\boldsymbol{\theta}_i$ and
$\boldsymbol{\theta}_j$ on the sky) can also be measured with
unprecedented accuracy \citep{Liao17}. We stress that lensing time delays are not affected by GW
damping effect due to DM viscosity. This advantage of strongly lensed
GW signals has been widely discussed in fundamental physics
\citep{Collett17} and cosmology \citep{Cao19}. Now, the angular
diameter distance to the lens can be calculated as
\begin{equation}
D_A(z_l)=D_{\mathrm{\Delta t}}\frac{D_{A}(z_l,z_s)}{D_{A}(z_s)}.
\end{equation}
We note that based on the Etherington reciprocity
relation, which is of fundamental importance in modern cosmology
\citep{Etherington33}, combined with distance sum rule, the luminosity distance to the source can be
expressed as
\begin{eqnarray}  \nonumber 
&& D_L(z_s) = \nonumber \\
 && = (1+z_s)(1+z_l)D_{\mathrm{\Delta
t}}\frac{D_{A}(z_l,z_s)/D_{A}(z_s)}{1- \left(D_{A}(z_l,z_s)/D_{A}(z_s)\right) (1+z_s)^{-2,}}
\nonumber
\end{eqnarray}
which is unaffected by GW damping effect and DM SIs.
Thirdly, another advantage of strongly lensed GW events comes from
their ability to provide simultaneous measurements of the effective
luminosity distance to the source, that is, coalescing binary NSs can be
calibrated as standard sirens. However, due to the magnification
effect of lensing, the luminosity distance measured from the
strongly lensed GW signals is not the true $D_{L,eff}$, but rather $D_{L,eff}/A$, where $A$ is the lensing
amplification factor. If the lensing magnification $\mu$ could be
independently measured through photometric EM observations,
we would be able to correct the amplitude of the GW signal for the amplification factor $A=\sqrt{\mu}$. \emph{\textup{Therefore, the true effective
luminosity distance from the observer to the source, which is
affected by viscous DM damping, could be derived as
}\begin{equation}\label{deff2}
D_{L,eff}(z_s, \beta)=D_L(z_s) e^{\beta
D_L(z_s)/2(1+z_s)^2}/\sqrt{\mu},\end{equation}
\textup{based on a detailed analysis of the images of the host galaxy and
the accompanying EM counterpart.}}

The function $\beta(z_l, z_s)$ can now be directly
obtained through the combined analysis with Eqs.~(2)-(5) for
individual lenses. This general function is applicable to any form
of SIDM along the line of sight. 
In the following section, we
apply this method to the simulated data of future
ground-based and space-based GW detectors.

\section{Simulated data and constraints}

A particular single strong-lensing system possesses its own sensitivity to DM viscosity. The recent analysis of \citet{Cutler09} revealed that the
DECIGO, a proposed space-based GW detector, can discover up to 1000
lensed GW events in the accumulated data of five to ten years. This is an improvement of more than one order of magnitude over the three nested ET
interferometers in the redshift range $z\leq5.00$ \citep{Li18}.
Following the approach proposed by \citet{Collett15}, we have
simulated a realistic population of strongly lensed GWs (10 and 1000 events in ET and in DECIGO, respectively), the
sampling distribution of which follows the intrinsic merger rate of
double compact objects (DCO) calibrated by strong-lensing effects
\citep{Ding15}. The lensed GW systems are randomly sampled on the sky, with the masses of the NSs uniformly sampled in the interval of $[1M_\odot,2M_\odot]$. The number density of lenses, assumed to have a Gaussian distribution of the central velocity dispersion similar to the SL2S sample ($\sigma_{0}=210\pm50$ km/s) \citep{Sonnenfeld13}, is characterized
by the velocity dispersion function (VDF) of early-type galaxies
from SDSS DR5 \citep{Choi07}. Based on the uncertainties of multiple measurements in strongly lensed gravitational waves (see the appendix for details), we numerically propagated the uncertainties of different observables (time delays, mass density profile, Einstein radius, velocity dispersion, line-of-sight contamination, etc.) to the uncertainty of different distances, and finally to the uncertainties of the DM
viscosity on which it depends.

Following the method outlined above, we show the
measurements of $\beta(z_l, z_s)$ expected to be detected by future
ET and DECIGO in Fig.~1. Our simulation is based on the
concordance $\Lambda$CDM model (collisionless CDM) with the parameters determined by Planck 2018 data \citep{Ade16}.

\emph{\textup{The question now is whether these measurements are sufficient to trace the distribution of DM and its properties.}} We find that
both ET and DECIGO would provide precise 3D measurements of the GW
damping rate $\beta(z)$ and DM viscosity $\eta(z)$ at different
redshifts and different positions on the sky. The significant discrepancy
between these individual measurements could be a probe of the large-scale distribution of DM and its properties.
In particular, the second-generation space-based
detectors (e.g. DECIGO) would provide a catalog of more precise $\beta(z)$ measurements that would be larger by one order of magnitude. This would increase the
possibility of determining different $\beta(z_l, z_s)$ for different
pairs $(z_l, z_s)$.
Summarizing the multiple $\beta$ measurements through the inverse-variance weighting, a constraint on the $\beta$
parameter at the precision of $\Delta \beta=10^{-4}\,{\rm
Mpc}^{-1}$ might be expected, with ten multi-messenger strongly lensed GW events from
ET. For comparison, we also considered the constraining power of the
same technique in space-borne detectors, using simulated data
representative of the future observations of lensed GW and EM
signals. The result is shown in Fig.~3. Lensed systems seen jointly
in GW and EM signals by DECIGO give much more stringent constraints,
the uncertainty of the damping of GWs in a viscous
Universe being $\Delta \beta=10^{-6}\,{\rm Mpc}^{-1}$. This can be
understood because the number of strongly lensed standard
sirens as well as of precise distance measurements at much higher
redshifts is higher. We note that statistically significant differences
between particular strong-lensing systems can be used to
distinguish between DM and modified gravity. Namely, as discussed for instance in some classes of modified gravity theories \citep{Belgacem2018a,Belgacem2018b,Mastrogiovanni2020}, GW
propagation involves a friction term. In consequence, GW amplitude
is attenuated in a similar manner as by DM shear viscosity. However,
this effect should be isotropic, while differences in DM
distribution along the line of sight could manifest themselves as
discussed in this section.

\begin{figure}
\begin{center}
\includegraphics[scale=0.4]{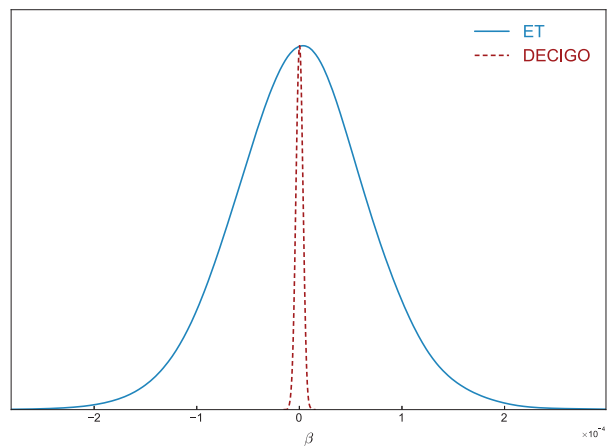}
\end{center}
\caption{Fits on the damping rate of GWs based on strongly lensed NS-NS systems from ET and DECIGO.}\label{fig2}
\end{figure}

\begin{figure}
\begin{center}
\includegraphics[scale=0.45]{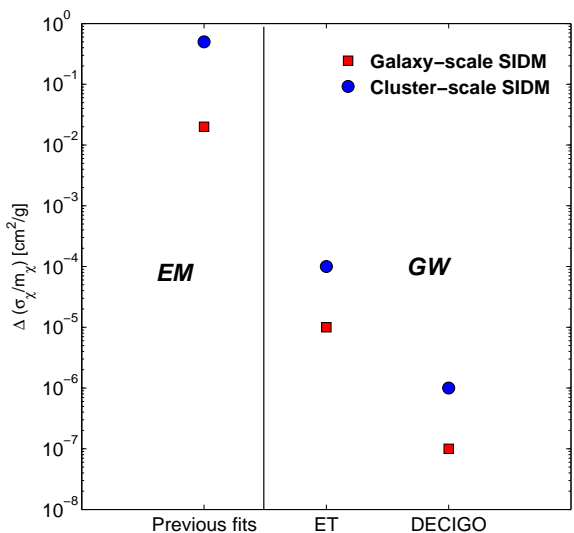}
\end{center}
\caption{Limits on the DM scattering cross-section per unit mass ($\sigma_{\chi}/m_{\chi}$) on the scales of galaxy and galaxy clusters, based on the current and future observations in the EM and GW domain.}\label{fig3}
\end{figure}

Now the main question now is whether\textup{\textup{\textit{\textup{ this accuracy is enough to
distinguish between possible DM shear viscosity and
SI cross-sections necessary to solve CCDM problems at
galaxy and cluster scales.}}}} SIDM is
parametrized by the scattering cross-section per unit mass
($\sigma_{\chi}/m_{\chi}$), which is generally a function of the
relative velocity of the DM particles {\bf{($\chi$)}}
\citep{Tulin18,Atreya},
\begin{eqnarray}\label{Eq:DM_beta}
\frac{\left \langle \sigma_{\chi} \right
\rangle}{m_{\chi}}=\frac{6.3\pi G\left \langle v \right
\rangle}{c^3 \beta}.
\end{eqnarray}
In particular, because the collision rate becomes negligible in
low-density regions, SIDM particles behave like collisionless CDM
at larger length scales. These newly proposed velocity-dependent
SIs provide a consistent fit to the CDM paradigm at
different scales, provided $\sigma_{\chi}/m_{\chi} \sim 1$ cm$^2$/g
on the scale of galaxies \citep{Dooley16} and $\sigma_{\chi}/m_{\chi}
\sim 0.1$ cm$^2$/g on the scale of galaxy clusters
\citep{Markevitch04}. However, the class of velocity-dependent SIDM
models remains largely unconstrained. Similarly, in this paper, two
SIDM scenarios are also considered with our
method, with the phenomenological SIDM halo model proposed in
\citet{{Kaplinghat2016}}. More specifically, focusing on
galactic-scale DM halos, ten lensed GWs detected by ET will
provide a stringent limit on its SI cross section as
\begin{equation}
\Delta(\sigma_{\chi}/m_{\chi}) \sim 10^{-4} cm^2/g.
\end{equation}
A much stronger limit on the DM SI cross section,
\begin{equation}
\Delta(\sigma_{\chi}/m_{\chi}) \sim 10^{-6} cm^2/g,
\end{equation}
can be obtained with 1000 lensed GW events detected by DECIGO.
Previously, by analyzing the rotation curves of five dwarf galaxies
in the THINGS sample \citep{Dh11} and seven LSB galaxies from
\citet{Kuzio08}, \citet{Kaplinghat2016} derived a limit on the cross
section $\Delta(\sigma_{\chi}/m_{\chi}) \sim 5\times10^{-1}$
cm$^2$/g for all 12 galaxies. In the present paper, using
10-1000 strongly lensed GWs from the third-generation ground-based
GW detectors and space-based detectors, our constraints are
$10^3\sim10^5$ times tighter than the results obtained before.
For the cluster-scale case, the limits we derive here are
$\Delta(\sigma_{\chi}/m_{\chi}) \sim 10^{-5}$ cm$^2$/g and
$\Delta(\sigma_{\chi}/m_{\chi}) \sim 10^{-7}$ cm$^2$/g in the
framework of ET and DECIGO, respectively. Therefore, we could obtain
very optimistic limits on the cluster-scale DM SI cross
section, which represents an improvement of three to five orders of
magnitude over the results using six clusters of galaxies
($\Delta(\sigma_{\chi}/m_{\chi}) \sim 2\times10^{-2}$ cm$^2$/g)
\citep{Kaplinghat2016}. The results are shown in Fig.~3. We
now briefly describe the effectiveness of our method in the context
of SIDM particle physics. To elaborate this idea, we take the
simplest SIDM model, which could be a real scalar field $s$ with quartic
SIs ($\lambda_s s^4/4!$)~\citep{deMartino20}. Although there are
several ways to build this type of models, the self-coupling
strength in the dark sector is characterized by $\lambda_s$, which
should be in the range of ($0<\lambda_s<\sqrt{\pi/9}$) due to
the stability of the potential and perturbation
limits~\citep{Cheng19}. Fig.~4 shows the preferred region of
the mass of the DM particle $m_s$ and the self-coupling
strength $\lambda_s$ on the scales of galaxies and galaxy clusters,
based on the simulated data regarding  strongly lensed GW signals
from the ET. Our final results demonstrate that in the framework of
SIDM capable of solving the small-scale issues of the CDM paradigm,
the DM particle mass (within some specific SIDM models) can be
assessed with an accuracy of $\sim O(0.1) \rm{GeV}$. This
conclusion, which could be extended to space-based detectors
(DECIGO) and other SIDM models (e.g. fermionic DM particles
that self-interact via a light mediator) \citep{Tulin13a,Tulin13b},
will strengthen the constraining power of our method to inspire new
observational programs and theoretical work in the future.

\begin{figure}
\begin{center}
\includegraphics[scale=0.45]{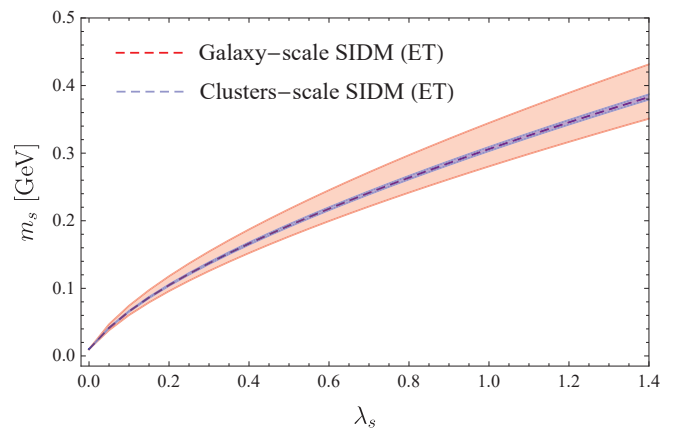}
\end{center}
\caption{Limits on the mass of DM particles and the
coupling constant reflecting the strength of SIs on
the scales of galaxies and galaxy clusters, based on lensed GW
signals in ET. The dotted lines denotes the central value, and the
68.3\% C.L. is shown by the red (blue) shaded region. }\label{fig4}
\end{figure}

\section{Summary and outlook}

The impact of the results presented in this work is expected to
increase significantly as more strongly lensed GWs are observed.
However, based on our results and arguments, the final question still
is whether\textit{\textup{ it is technically possible to achieve uncertainties this small for the measurements of DM self-scattering.}} Many
technical problems would be involved from the observational point of
view. For instance, the current low detection rate of inspiraling
double compact object detection and lensing rate may be due to the
lack of either the necessary high temporal resolution or a wide
enough field of view in the current EM and GW detectors.
Fortunately, the upcoming and planned GW detectors (e.g. ET and
DECIGO) and wide-area and deep surveys (e.g. the LSST), with a much
wider field of view and higher sensitivity, will be able to discover
and precisely localize a large number of strongly lensed GW and EM
events. However, it is very likely that only some fraction of the GWs with EM counterparts will be measured. This means that we would observe some lensed GW signals from NS-NS systems without an electromagnetic counterpart. Moreover, accurate lens reconstruction could also be difficult when the Einstein radius of the lens is too small or the background galaxy is too dim for a  redshift measurement. Therefore, we need to be cautious about the number of strongly lensed GW signals that allow both the determination of EM counterparts and precise lens reconstruction. Secondly, the systematic floor needs to be further studied for future works in order to let us know how powerful this lensing method might be in probing the distribution of DM exhibiting elastic and velocity-independent SIs. For instance, we still need precise measurements of the image magnification (or the amplification factor of GW signals), which calls for more knowledge about the AGN accretion models and local image environments from astrophysics inputs. Finally, if the estimates of the mean collision velocity for DM particles have large uncertainties, the results shown in Fig.~3-4 might be affected. Accounting for the uncertainty of the mean collision velocity on the scales of galaxy and cluster, Eq.~(6) might make these limits less stringent. To summarize, with more abundant observational information in the future, we will have a better understanding of the SI nature of DM at both galaxy and cluster scales, and we will also know better how
to use this as a probe of the large-scale distribution of DM in the GW domain, as discussed in this Letter.

\begin{acknowledgements}

This work was supported by the National Natural Science Foundation of China under Grants Nos. 12021003, 11690023, and 11920101003; the National Key R\&D Program of China No. 2017YFA0402600; the Strategic Priority Research Program of the Chinese Academy of Sciences, Grant No. XDB23000000; the Interdiscipline Research Funds of Beijing Normal University; the China Manned Space Project (Nos. CMS-CSST-2021-B01 and CMS-CSST-2021-A01); and the CAS Project for Young Scientists in Basic Research under Grant No. YSBR-006. M.B. was supported by Foreign Talent Introducing Project and Special Fund Support of Foreign Knowledge Introducing Project in China (No. G2021111001L).

\end{acknowledgements}

\section*{Appendix}

We summarize the uncertainties of multiple measurements in strongly lensed NS-NS systems, that is, the precision of lens reconstruction, the precision of time delays, and the precision of lensed standard sirens. The error strategy below is used to produce the measurements of $\beta(z_l, z_s)$ expected to be derived by future GW detectors (see Fig.~1 for details).

\subsection*{Precision of lens reconstruction}

Three sources of uncertainty are included in our simulation of lens reconstruction.
With the observation of host arcs by deep imaging of the HST, the
Einstein radius of multiple images and the lens mass slope can be
measured precisely and accurately (at the level of 1\% ) based on advanced lens modelling techniques \citep{Suyu10,Suyu12} and
kinematic modelling methods \citep{Auger10,Sonnenfeld12}. This error
strategy has been extensively used in the simulation of the LSST lens
sample, with high-quality (sub-arcsecond) imaging data in general
\citep{Collett16}. Although the joint lensing and
stellar-dynamical analysis could generate 5\% accuracy on the total
mass-density slope \citep{Ruff11}, the lessons learnen from lensed
quasar results showed that this accuracy is expected to increase to 1\%, with supplementary information about time delays
obtained by the COSMOGRAIL \citep{Bonvin18} or H0LiCOW collaboration
\citep{Suyu17}. Meanwhile, we take 5\% as the fractional uncertainty
of the observed velocity dispersion, with detailed follow-up
spectroscopic information from other ground-based facilities, that is,
multi-object and single-object spectroscopy to enhance dark energy
science from LSST \citep{Hlozek19}.

\subsection*{Precision of time delays}

We consider three sources of
uncertainty in the simulation of time-delay measurements. It is well
recognized that coalescing NS-NS systems manifested as short
gamma-ray bursts can have many advantages as precise lensing time-delay indicators. However, lensed GW signals  could provide
extremely accurate $\Delta t$ measurements, with a detailed analysis
of the waveform from the GW detection pipeline \citep{Liao17}
(accuracy ranging from $10^{-4} \; ms$ to $0.1\;s$). Therefore, the
fractional uncertainty of $\Delta t$ could be neglected. On the
other hand, the detection of the EM counterpart would be very
beneficial. The host galaxy image followed up with a quality typical
to HST observations with dedicated lens modelling
techniques would result in a precision of 0.5\% in lens modelling;
specifically, the Fermat potential difference reconstruction for a
well-measured lensed GW system \citep{Liao17}. The
analogous uncertainty is $3\%$ for almost all quasar-galaxy systems,
with lensed quasar images quality typical to HST observations \citep{Suyu17}. In this analysis, we applied a strategy to calculate the uncertainty of $\Delta \phi$, from the measurements of the lens mass profile, the Einstein radius, and
their corresponding uncertainties \citep{Cao19}. Finally, the effect
of the light-of-sight density fluctuation on the gravitational time
delays should be taken into account. For each lens system, we added an
additional 1\% relative uncertainty to the lens modelling, which
naturally appears due to mass along the line of sight to the source
\citep{Seljak94}. This additional line-of-sight systematic uncertainty
(at the 1\% level) is also characteristic for that derived in typical strongly lensed quasar systems \citep{Suyu17}.

\subsection*{Precision of lensed standard sirens}

We account for two sources of uncertainties in the simulation of lensed GWs: effective luminosity distance, and lensing magnification. Firstly, to obtain the uncertainty of $D_{L,eff}(z)$ from the GW  signal of NS-NS standard
sirens observed by different detectors, the distance precision was
taken as $\sigma^2_{\rm tot}= \sigma^{2}_{\rm ins}+ \sigma^{2}_{\rm
len}$, with the instrumental measurement uncertainty
$\sigma_{inst}\simeq 2D^L_s/\rho_{net}$, where $\rho_{net}$ denotes
the combined signal-to-noise ratio (S/N) for the network of
independent interferometers, that is, the square root of the inner
product of the strain in Fourier space \citep{Cai15}. A detailed description of the noise power spectral density (PSD), which characterizes the sensitivity of the detector, could be found in \citet{Cai15} and \citet{Kawamura06}. Note that we assumed an ideal situation by setting the inclination angle of the binary orbital angular momentum $\iota=0$. However, \cite{Nissanke10} demonstrated the strong correlation between the luminosity distance and the inclination angle of the binary ($\iota=[0, 180^\circ]$). This distance-inclination degeneracy, which decreases the sensitivity to both parameters, might be broken by the strongly beamed observations ($\iota<20^\circ$) of short gamma-ray bursts as the electromagnetic counterparts of GWs \citep{li2015extracting}. Therefore, considering the maximum effect of the inclination on the combined S/N, we chose to double the instrumental uncertainty of $D_L$ as the upper limit, as was proposed in the recent study of the third-generation GW detector (ET) based on a Fisher information matrix \citep{cai2017estimating}. This strategy has been extensively applied in subsequent works focusing on the constraint ability of standard siren GWs on different cosmological parameters \citep{Geng20,Pan21,Zheng21,Cao22}. Moreover, the measurements of luminosity distance are also affected by the weak-lensing effect, which is modeled as $\sigma_{lens}/D^L_s=0.05z$ for ET \citep{Zhao11} and $\sigma_{lens}/D^L_s=0.044z$ for DECIGO \citep{Cutler09}. Secondly, the amplification factor of the GW signal along with the corresponding uncertainty is related to the magnification ($\mu$) and its uncertainty of the EM counterpart, following the recent discussion of gravitational lensing statistics and the magnification by galaxies \citep{Suyu10,Suyu12}. More specifically, the uncertainty of $\mu$ is calculated from the observables and their uncertainties (i.e. the Einstein radius and the mass profile slope) \citep{Cao19} by solving the lens equation with \emph{glafic} in the framework of a specific lens model \citep{Oguri10,Oguri10b}. In our simulated data, the image magnification is expected to be determined at a level of $\sim10\%$. Meanwhile, recent works have studied the microlensing (ML) effect generated by stars in lensing galaxies, which potentially biases the magnification map of strongly lensed sources \citep{Foxley-Marrable18}. Although the community is mainly of the opinion that this effect needs to be incorporated in a strong-lensing analysis, much uncertain input for the ML priors on the AGN accretion disk models remains \citep{Dexter11}, as well as on the local environments for images and mass function for the stars \citep{Chen18}. Therefore, to assess the impact of ML on luminosity distance estimates, we assumed an additional systematic 5\% uncertainty on the magnification measurements for all lensed gravitational waves. A similar error strategy can be found in the recent discussion of \citet{Liao19}, focusing on ML time delays from the light curves of strongly lensed quasars. We emphasize that for galactic-scale lenses with point-like background sources, the model predictions in the generic case for image magnifications could be very large \citep{Yahalomi17}. However, for strongly lensed standard sirens in the GW domain, the lensing magnification could be more precisely measured in a lens-model-independent way by comparing the observed flux with that of other unlensed GWs within a narrow redshift bin \citep{Goobar17}.

\end{document}